\documentclass[a4paper,12pt]{article}
\usepackage{feynmp-auto,expdlist}
\usepackage{amsmath, amsfonts}
\usepackage{graphicx,arydshln}
\usepackage{enumerate}
\usepackage{hyperref}
\usepackage{latexsym}
\usepackage{dsfont}
\usepackage{hepnicenames}
\usepackage{enumerate}
\usepackage{soul}
\usepackage[normalem]{ulem}
\usepackage{comment}

\newcommand{\footnoten}[1]{}

\usepackage[font={small},textfont={it}]{caption} 

\newcommand{\nnn}[1]{{\color{verdes}#1}}
\renewcommand{\nnn}[1]{#1}

\def\rt{{\tilde{r}}}

\def\rt{{{r}}}

\usepackage{units}

\usepackage{mathrsfs,graphicx,rotating,amsmath,amsfonts,mathtools,booktabs,amssymb,wasysym}
\usepackage{hyperref}\usepackage{slashed}
\usepackage[nosort]{cite}
\usepackage[table,xcdraw,dvipsnames]{xcolor}
\usepackage{bm}
\usepackage{multirow,multicol}
\hypersetup{colorlinks,bookmarksopen,bookmarksnumbered,
	linkcolor=blus,pdfstartview=FitH,urlcolor=rossos,citecolor=verde}
\allowdisplaybreaks

\newcommand{\myfootnote}[1]{}
\newcommand{\myomit}[1]{{\color{gray}#1}}
\renewcommand{\myomit}[1]{}

\renewcommand{\[}{\left[}

\newcommand{\bP}{\bar{M}_{\rm Pl}}

\def\Lag{\mathscr{L}}

\newcommand{\mio}[1]{}

\newcommand{\med}[1]{\langle #1\rangle}

\def\bpm{\begin{pmatrix}}
	\def\epm{\end{pmatrix}}

\usepackage{mathrsfs}

\newcommand{\fig}[1]{~\ref{fig:#1}}
\newcommand{\sfrac}[2]{#1/#2}

\renewcommand{\Im}{{\rm Im}\,}
\allowdisplaybreaks
\usepackage{multicol}
\usepackage{color}
\definecolor{rosso}{cmyk}{0,1,1,0.4}
\definecolor{rossos}{cmyk}{0,1,1,0.55}
\definecolor{rossoc}{cmyk}{0,1,1,0.2}
\definecolor{blu}{cmyk}{1,1,0,0.3}
\definecolor{blus}{cmyk}{1,1,0,0.6}
\definecolor{bluc}{cmyk}{1,1,0,0.1}
\definecolor{verde}{cmyk}{0.92,0,0.59,0.25}
\definecolor{verdec}{cmyk}{0.92,0,0.59,0.15}
\definecolor{verdes}{cmyk}{0.92,0,0.59,0.4}

\newcommand{\bp}{\bar{M}_{\rm Pl}}
\newcommand{\eq}[1]{~{\rm (\ref{eq:#1})}}

\newcommand{\MeV}{\,{\rm MeV}}
\newcommand{\GeV}{\,{\rm GeV}}
\newcommand{\TeV}{\,{\rm TeV}}

\def\circa#1{\,\raise.3ex\hbox{$#1$\kern-.75em\lower1ex\hbox{$\sim$}}\,}

\newcommand{\beq}{\begin{equation}}
\newcommand{\eeq}{\end{equation}}

\newcommand{\bea}{\begin{eqnarray}}
\newcommand{\eea}{\end{eqnarray}}
\newcommand{\be}{\begin{equation}}
\newcommand{\ee}{\end{equation}}
\font\tenrsfs=rsfs10 at 12pt
\font\sevenrsfs=rsfs7
\font\fiversfs=rsfs5
\newfam\rsfsfam
\textfont\rsfsfam=\tenrsfs
\scriptfont\rsfsfam=\sevenrsfs
\scriptscriptfont\rsfsfam=\fiversfs

\newsavebox\MBox

\renewenvironment{thebibliography}[1]
{\begin{multicols}{2}[\section*{\refname}]%
		\@mkboth{\MakeUppercase\refname}{\MakeUppercase\refname}%
		\list{\@biblabel{\@arabic\c@enumiv}}%
		{\settowidth\labelwidth{\@biblabel{#1}}%
			\leftmargin\labelwidth
			\advance\leftmargin\labelsep
			\@openbib@code
			\usecounter{enumiv}%
			\let\p@enumiv\@empty
			\renewcommand\theenumiv{\@arabic\c@enumiv}}%
		\sloppy
		\clubpenalty4000
		\@clubpenalty \clubpenalty
		\widowpenalty4000%
		\sfcode`\.\@m}
	{\def\@noitemerr
		{\@latex@warning{Empty `thebibliography' environment}}%
		\endlist\end{multicols}}

\newcommand{\kg}{\,{\rm kg}}

\renewcommand{\L}\Lag

\def\circa#1{\,\raise.3ex\hbox{$#1$\kern-.75em\lower1ex\hbox{$\sim$}}\,}
\makeatletter

\font\ital=cmu10

\def\hhref#1{\href{http://arxiv.org/abs/#1}{arXiv:#1}}
\usepackage{xstring}
\newcommand{\hhrefq}[1]{\IfSubStr{#1}{:}{\href{http://inspirehep.net/search?ln=en&ln=en&p=#1&of=hb&action_search=Search&sf=&so=d&rm=&rg=25&sc=0}{InSpire:#1}}{\hhref{#1}}}

\def\art{\@ifnextchar[{\eart}{\oart}}
\def\eart[#1]#2#3#4#5#6{{\rm #2}, {\em #3 \bf #4} {\rm (#6) #5} ({\em #1})}
\def\article{\@ifnextchar[{\earticle}{\oarticle}}
\def\oarticle#1#2#3#4#5#6{{\rm #1}, {\ital `#6'}, {\rm #2 #3 (#5) #4}}
\def\earticle[#1]#2#3#4#5#6#7{{\rm #2}, {\ital `#7'}, {\rm #3 #4 (#6) #5}  [\hhrefq{#1}]}
\def\hepart[#1]#2{{\rm #2, \sl#1}}
\def\heparticle[#1]#2#3{#2, {\ital `#3'} [\hhrefq{#1}]}
\newcommand{\doi}[1]{\href{http://dx.doi.org/#1}{[link]}}

\newcommand{\hhrefqq}[1]{\IfBeginWith{#1}{10.}{\href{https://doi.org/#1}{doi:#1}}{\hhrefq{#1}}}
\def\earticle[#1]#2#3#4#5#6#7{{\rm #2}, {\ital `#7'}, {\rm #3 #4 (#6) #5}  [\hhrefqq{#1}]}

\renewenvironment{thebibliography}[1]
{\begin{multicols}{2}[\section*{\refname}]%
		\@mkboth{\MakeUppercase\refname}{\MakeUppercase\refname}%
		\list{\@biblabel{\@arabic\c@enumiv}}%
		{\settowidth\labelwidth{\@biblabel{#1}}%
			\leftmargin\labelwidth
			\advance\leftmargin\labelsep
			\@openbib@code
			\usecounter{enumiv}%
			\let\p@enumiv\@empty
			\renewcommand\theenumiv{\@arabic\c@enumiv}}%
		\sloppy
		\clubpenalty4000
		\@clubpenalty \clubpenalty
		\widowpenalty4000%
		\sfcode`\.\@m}
	{\def\@noitemerr
		{\@latex@warning{Empty `thebibliography' environment}}%
		\endlist\end{multicols}}

\newcounter{alphaequation}[equation]
\def\thealphaequation{\theequation\hbox to
	0.6em{\hfil\alph{alphaequation}\hfil}}
\def\eqnsystem#1{
	\def\@eqnnum{{\rm (\thealphaequation)}}
	\def\@@eqncr{\let\@tempa\relax \ifcase\@eqcnt \def\@tempa{& & &} \or
		\def\@tempa{& &}\or \def\@tempa{&}\fi\@tempa
		\if@eqnsw\@eqnnum\refstepcounter{alphaequation}\fi
		\global\@eqnswtrue\global\@eqcnt=0\cr}
	\refstepcounter{equation} \let\@currentlabel\theequation \def\@tempb{#1}
	\ifx\@tempb\empty\else\label{#1}\fi
	\refstepcounter{alphaequation}
	\let\@currentlabel\thealphaequation
	\global\@eqnswtrue\global\@eqcnt=0 \tabskip\@centering\let\\=\@eqncr
	$$\halign to \displaywidth\bgroup \@eqnsel\hskip\@centering
	$\displaystyle\tabskip\z@{##}$&\global\@eqcnt\@ne
	\hskip2\arraycolsep\hfil${##}$\hfil& \global\@eqcnt\tw@\hskip2\arraycolsep
	$\displaystyle\tabskip\z@{##}$\hfil
	\tabskip\@centering&\llap{##}\tabskip\z@\cr}
\def\endeqnsystem{\@@eqncr\egroup$$\global\@ignoretrue} \makeatother

\definecolor{Gray}{gray}{0.95}

\def\bal#1\eal{\begin{align}#1\end{align}}

\newcommand{\bk}[2]{\langle #1  |  #2  \rangle}

\newcommand{\bAk}[3]{\langle #1  |  #2|#3  \rangle}

\begin{document}
\thispagestyle{empty}
\begin{center}
{\huge \bf \color{rossos} Triggering Higgs vacuum decay}  \\[3ex]
{\bf\large Alessandro Strumia}  \\[1ex]
{\it  Dipartimento di Fisica dell'Universit{\`a} di Pisa}\\[3ex]
{\large\bf\color{blus} Abstract}
\begin{quote}
The Standard Model Higgs potential seems unstable at field values $h > h_{\rm top} \sim 10^{10}\GeV$.
Vacuum decay can be triggered by $N\sim 4\pi/\lambda \sim 1000$ overlapped Higgs bosons 
with energy $\sqrt{\lambda} h_{\rm top}$.
However, this configuration is stimulated by ultra-high energy collisions with a $\exp(-{\cal O}(N))$ suppression,
comparable to spontaneous vacuum decay: no `Higgspolosion' enhancement arises.
This implies that ultra-high energy cosmic ray collisions are safe, despite that their number (in production sites)
likely is tens of orders of magnitude higher than what usually estimated (in space).
We speculate on how vacuum decay could be induced classically
at futuristic ultra-high energy colliders (either using a few right-handed electron beams,
or a double collision producing an in-coming wave of $N$ boosted Higgs bosons)
and on how the resulting vacuum bubble could be controlled to extract energy.
\end{quote}
\end{center}

\tableofcontents
\normalsize


\section{Introduction}
The discovery of Higgs boson $h$ opens the door to a new form of energy, the vacuum energy in the Higgs potential $V(h)$.
The Higgs currently sits at the Standard Model (SM) local minimum of its potential.
If this is the global minimum, no vacuum energy remains stored in our Universe, and vacuum cannot be used as a source of energy.
However, extrapolating the SM Higgs potential $V(h)$ to ultra-high field values $h$
suggests that the potential might have a local maximum at $h=h_{\rm top}\sim10^{10}\GeV$ becoming negative 
at $h > e^{1/4} h_{\rm top}$~\cite{1307.3536,1505.04825}.
This happens for current best-fit values of the top quark mass, Higgs mass and strong coupling.
Within current uncertainties a much larger instability scale or even stability is possible.
Clarifying if $V(h)$ is really unstable needs improved future measurements of the top mass, doable at a lepton collider at the top threshold
(such as an $e^- e^+$ collider in the LEP tunnel, or a $\mu^-\mu^+$ collider)~\cite{2203.17197}.


\smallskip

Assuming that the Higgs instability exists,
we explore under which conditions it can result in vacuum decay.
The following possibilities have been considered in the literature:
\begin{itemize}
\item Spontaneous vacuum tunnelling~\cite{Coleman:1977py} is exponentially slow~\cite{hep-ph/0104016,0712.0242,1307.3536,1505.04825,1707.08124}.
\item The thermal energy in the early universe at temperatures $T \circa{>} h_{\rm top}$ could have overcome the potential barrier in $V(h)$, 
but thermal effects also add a thermal barrier:
as a result vacuum decay remains exponentially slow at any $T$ (see e.g.~\cite{Arnold:1991cv}).

\item Inflationary fluctuations could have trigger the instability, but this depends on the unknown inflation model and non-minimal Higgs coupling to gravity
(see e.g.~\cite{1505.04825}).

\item Small hypothetical primordial black holes with Hawking temperature $T \circa{>} h_{\rm top}$ are a possible seed of vacuum decay,
but the rate is again exponentially slow, similarly to thermal vacuum decay
(claims in the literature differ, I think this is the correct conclusion as discussed in~\cite{2111.08017,2209.05504}).

\item Vacuum decay seeded by particle collisions with ultra-high energy $E \circa{>} h_{\rm top}$ has been considered for some scalar potentials,
finding that the rate remains partially exponentially suppressed~\cite{Affleck:1979px,Selivanov:1985vt,Kiselev:1992hk,Ellis:1990bv,hep-ph/9309237,hep-ph/9704431,hep-th/0412253,hep-th/0511095,1004.2015,1503.06339,1509.05405}. 
\end{itemize}
In this paper we reconsider the latter issue.
The exponential suppression persists because, as discussed in section~\ref{trig}, 
 the minimal configuration needed to seed vacuum decay contains $N \gg 1$ scalar quanta.
In the case of SM Higgs vacuum decay, one needs $N \sim 4\pi/\lambda \sim 1000$ Higgs boson quanta.
Such a semi-classical state is generated out of collisions of a few quanta with rate exponentially suppressed by a `one to many particles' factor.
In section~\ref{bubble} we revisit and extend previous computations, focusing on the specific case of Higgs vacuum decay,
that needs going beyond the thin-wall approximation, and summarizing why 
 cross sections for producing $N$ Higgs bosons remain small at large $N$ (no `Higgsplosion' happens).
 
In section~\ref{CR} we reconsider collisions of two ultra-high-energy cosmic rays (CR), showing that a significantly enhanced CR collision rate happened
if cosmic rays are accelerated by relatively compact astrophysical objects, from magnetars up to Active Galactic Nuclei.
We will also find that few-particle ultra-high energy CR collisions can have happened, while many-particle collisions never occurred.
As a result, CR collisions could not trigger Higgs vacuum decay.

\medskip

The main simple lesson is that an appropriate many-particle classical process is needed to trigger
vacuum decay avoiding the exponential suppression that would make the rate negligible.
We  then explore in section~\ref{many}
if the critical configuration could be engineered classically via futuristic collider schemes with ultra-high energy $\sqrt{s}\circa{>} h_{\rm top}$.
We next discuss if the true vacuum bubble can be artificially controlled and used to extract energy from the vacuum.

Section~\ref{concl} presents our conclusions.



\section{Triggering Higgs vacuum decay}\label{trig}
We consider an initial configuration of the Higgs field $h(t,r)$ at $t=0$ 
with  field value $h\sim h_0$ within a space region of size $r\circa{<}r_0$.
For simplicity we assume spherical symmetry, denote as $r$ the radius, and assume
that the initial configuration is at rest. An example is
\beq \label{eq:hin}
h(0,r) = \frac{h_0}{1+r^2/r_0^2},\qquad \dot h(0,r)=0.\eeq 
The classical Higgs field equation in flat space is $\ddot h - h'' - 2h'/r = -V'$, 
where, as usual, $\dot h = \partial h/\partial t$ and $h' = \partial h/\partial r$.

\subsection{Simple estimates}
The total energy of the configuration is
\beq \label{eq:Eh0} E = \int dr\, 4\pi r^2 \left[\frac{h'^2}{2} +  V\right]  \sim 4\pi r_0 h_0^2  [1 + \lambda r_0^2  h_0^2]\eeq
having here simply approximated the potential energy as $V=\lambda h^4/4$.

In its subsequent evolution this field configuration triggers vacuum decay rather than dissolving if two conditions are met:
\begin{enumerate}
\item[1)] $h_0 \circa{>} h_{\rm top}$ is over the top of the SM potential barrier;
\item[2)] $h_0 r_0\sqrt{|\lambda|} \circa{>}1$ such that 
the potential energy wins over the gradient energy in the classical evolution.
This estimate can be obtained by demanding that potential energy dominates over the gradient energy in the total energy of eq.\eq{Eh0}, 
or by considering the classical equation of motion at the point $r=0$,
or  in thin-wall approximation (eq.\eq{Hbubble} later).
\end{enumerate}
According to conditions 1) and 2) above, an initially static configuration $h(0,r)$ evolves towards the true vacuum if its energy is
\beq E\circa{>} E_{\rm min}\sim \frac{4\pi h_{\rm top}}{\sqrt{|\lambda|}}.\eeq 
The Fourier transform of $h(0,r) $ is large at $k \circa{<} 1/r_0$, showing that the scalar field configuration
contains 
\beq N  \sim \frac{E}{k}  \sim 4\pi  (h_0 r_0)^2\circa{>} N_{\rm min} \sim \frac{E_{\rm min}}{k}  \sim \frac{4\pi }{|\lambda|}\eeq
Higgs quanta.

\subsection{Numerical computations}
We validate and precise the above analytic approximations by 
numerically solving the classical Higgs field equation in flat space.
The SM Higgs potential can be accurately approximated through an effective running coupling as
\beq\label{eq:VhSM}
V (h)\approx \lambda(h)\frac{h^4}{4} ,\qquad
\lambda(h) \approx -b \ln \frac{h^2}{e^{1/2}h_{\rm top}^2}\qquad\hbox{with}\qquad  b \approx \frac{0.15}{(4\pi)^2} 
\eeq
for best-fit values of the SM parameters.
Switching to the dimension-less field $\tilde h=h/h_{\rm top}$ and coordinates $\tilde x = x h_{\rm top}\sqrt{b}$, the Higgs action simplifies to
\beq S  = \frac{1}{b} \int d^4\tilde x \,  \left[\frac{1}{2}\left(\frac{\partial\tilde h}{\partial \tilde x_\mu}\right)^2+
\frac{\tilde h^4}{4}\ln \frac{\tilde{h}^2}{e^{1/2}}\right],\eeq
showing that the dynamics depends in a simple way on the values of $b$ and $h_{\rm top}$.

\smallskip

\begin{figure}[!t]
\begin{center}
\includegraphics[width=0.45\textwidth]{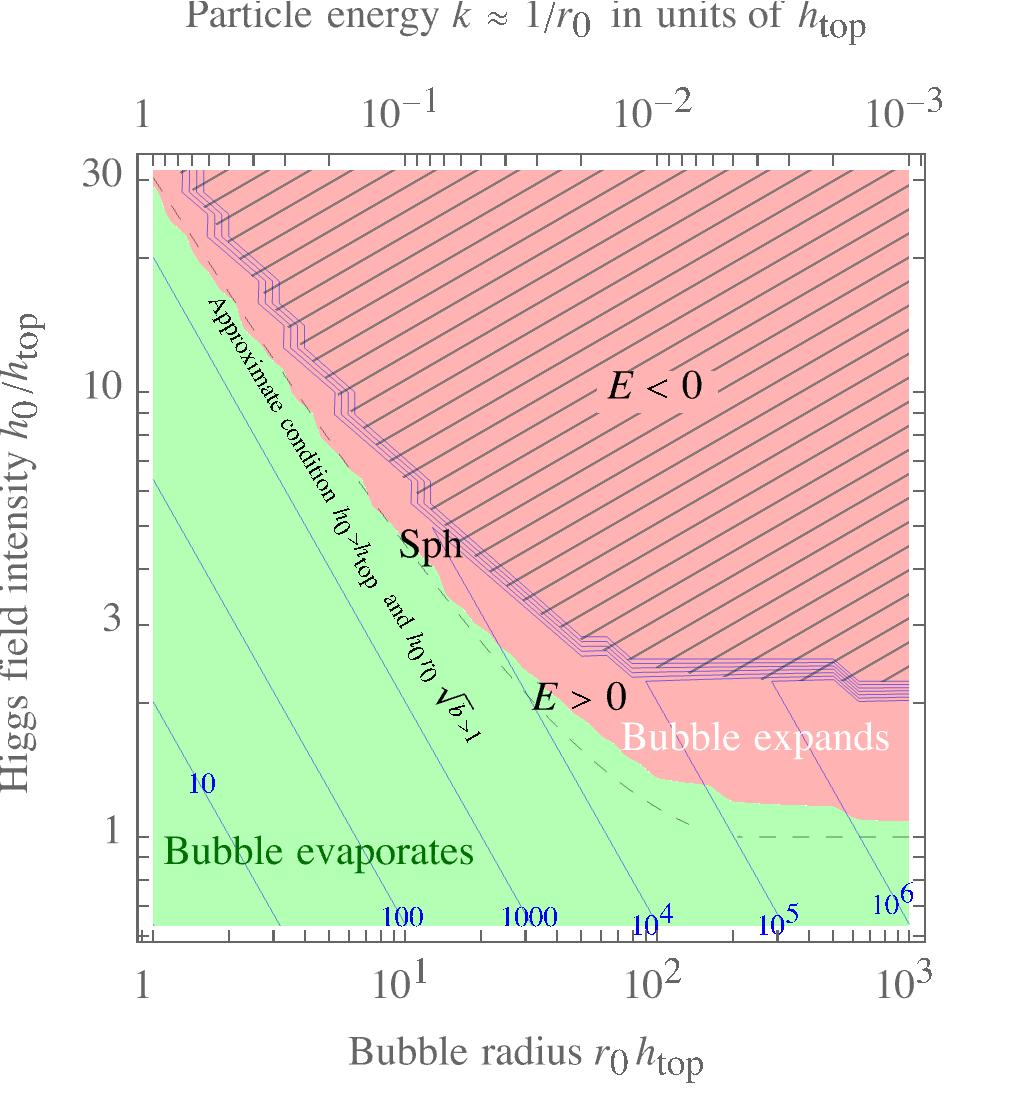}\qquad
\includegraphics[width=0.42\textwidth]{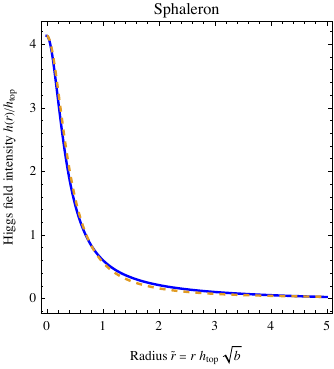}
\caption{\label{fig:dang}  {\bfseries Left}: 
We numerically evolve a spherical Higgs bubble $h_0/(1+r^2/r_0^2)$ initially at rest.
In the red region the bubble expands triggering vacuum decay,
while in the green region the bubble dissolves.
The red region is well approximated as $h_0 > h_{\rm top}$ and $h_0r_0 > 1/\sqrt{|\lambda|}$ (dashed curve).
The hatching covers the unphysical region where the bubble is so large that its  energy is negative, $E<0$.
The blue iso-contours show the number $N $ of Higgs quanta.
This confirms that expansion needs $N\circa{>}4\pi/|\lambda|$ quanta naively computed from the static configuration at $t=0$.
The sphaleron is denoted as `Sph'.
{\bfseries Right}: sphaleron solution, and its analytic approximation (dashed curve).}
\end{center}
\end{figure}

We evolve the Higgs field starting from  an initial field profile with vanishing time derivative.
The result for the profile in eq.\eq{hin} is shown in fig.\fig{dang}a:
the `drainage divide' critical boundary in the $(h_0, r_0)$ plane that separates initial configuration that 
end up in the false vacuum or in the true vacuum is well approximated by the analytic conditions 1) and 2).
Fig.\fig{dang}a also shows iso-curves of the approximate number of Higgs quanta, 
computed from the Fourier transform 
\beq   \label{eq:N}
h_k =  \int dr\, 4\pi r^2 \, h(r) \frac{\sin kr}{kr}
 \qquad\hbox{as}\qquad
N  = \int \frac{dk~4\pi k^2}{(2\pi)^3k} \frac{k^2}{2} |h_k|^2.\eeq
The expression for $N$ applies to an in-going or out-going Higgs wave with $h \ll h_{\rm top}$ such that Higgs quanta are free with energy $k$.
So eq.\eq{N}  can be applied to the field configuration at a time such that
the bubble is dissolving into the false vacuum, and field values $h$ got much below $h_{\rm top}$.\footnote{For simplicity, in fig.\fig{dang} we  evaluate eq.\eq{N}  at $t=0$ on the static configuration of eq.\eq{hin} with $h \sim h_{\rm top}$
obtaining $N=\left(\sfrac{\pi r_0 h_0}{2}\right)^2$. 
\nnn{This is a good approximation, as confirmed by a
dedicated analysis of various configurations: letting them evolve up to longer times and weaker field values
shows that true number of quanta is $k N/2 \approx N$}, 
where the factor $1/2$ accounts for the static initial configuration,
and $k$ for the potential energy, with $k\sim 1-2$ for critical bubbles.
\nnn{This simplification is avoided when computing incoming waves in fig.\fig{banghik} and\fig{banglowk}}.
}

\smallskip

Fig.\fig{dang}a shows that increasing the energy $k \sim 1/r_0$ of each quantum does not reduce the needed $N$  below 
$N_{\rm min}\approx 1/b\approx 1000$,
while $N \gg N_{\rm min}$ quanta are needed for a large bubble made of quanta with energy lower than the
optimal energy $k \sim \sqrt{|\lambda|} h_{\rm top} \sim 0.1 h_{\rm top}$.
A nearly identical figure is obtained replacing eq.\eq{hin} with $h(r)= h_0 e^{-r^2/r_0^2}$, for which $N=\pi(r_0 h_0)^2/2$ is mildly lower.

\smallskip

A special configuration is the sphaleron, the static spherical solution to the Higgs field equation.
Thereby it is the critical configuration with lowest energy
\beq \label{eq:Esph}
E_{\rm min}=E_{\rm sph}\approx \frac{9.7 h_{\rm top}}{\sqrt{b}} \approx  500\, \hbox{Joule} \frac{h_{\rm top}}{10^{10}\GeV}.\eeq
The sphaleron has $h_0 \approx 4.1 h_{\rm top}$ and its profile is plotted in fig.\fig{dang}b.
It can be approximated as eq.\eq{hin} with
$r_0\approx 0.4/h_{\rm top}\sqrt{b}$, corresponding to the point denoted as `Sph' in fig.\fig{dang}a.
By letting the sphaleron dissolve in time~\cite{Hellmund:1991ub}, one finds $N_{\rm sph} \approx 2.2/b \circa{>} N_{\rm min}$.
Critical configurations with higher energy $E > E_{\rm sph}$ do not have $N$ significantly lower than $N_{\rm sph}$.
This physically suggests that vacuum decay stimulated by particle collisions will remain exponentially suppressed, as discussed in the next section.




\section{Vacuum decay stimulated by few particles}\label{bubble}
We have seen that a Higgs field configuration that
evolves into the true vacuum contains $N > N_{\rm min}\sim 4\pi/|\lambda| \sim 10^3$ Higgs quanta.
Then, one expects that the probability of generating such a configuration out
of high-energy collisions with few particles is suppressed by an exponential factor $\sim e^{-{\cal O}(N)}$
parametrically similar to the vacuum tunnelling rate $\sim e^{-S_{\rm bounce}}$.

To see this, let us over-simplify the problem considering free Higgs particles, such that one can define
the number operator  $\hat N_k = \hat a^\dagger_k \hat a_k$ in terms of the creation operator  $\hat a^\dagger_k$.
As well known,  states with exactly $N$ quanta have zero field value $\bAk{N}{\hat h}{N}$ and large fluctuations.
The good semi-classical states with small field fluctuations and large $N=\langle \alpha | \hat N |\alpha\rangle = |\alpha|^2$
are instead the coherent states 
$ |\alpha\rangle = e^{-|\alpha|^2/2} \sum_{n=0}^\infty {\alpha^n} |n\rangle/{\sqrt{n!}}$,
eigenstates of $\hat{a} |\alpha\rangle = \alpha |\alpha\rangle$.
The scalar product between two coherent states  is
exponentially suppressed as $|\bk{\alpha}{\beta}|^2=e^{-|\alpha-\beta|^2}$.


\smallskip

A qualitatively similar suppression persists  taking into account the interaction in the scalar potential.
The vacuum decay rate induced by particle collisions was computed in thin-wall approximation by Voloshin~\cite{hep-ph/9309237}
using the following WKB-like result from Landau~\cite{Landau}.
Let us consider (for simplicity) a quantum system described by an Hamiltonian $H$ that depends on
one degree of freedom $q$ and its conjugated momentum $p$.
The matrix element of an operator ${\cal O}$ among quasi-classical states with energies $E_1$, $E_2$ is
not easily computed, because their wave functions oscillate wildly, averaging to nearly zero.
Suitable analytic continuations to complex $q$ in the direction of decreasing WKB exponentials show that
the matrix element is suppressed as~\cite{Landau}
\beq\label{eq:Landau}
\bAk{E_2}{{\cal O}}{E_1}\sim \exp\left[-{\rm Im} \left(  \int_{q_1}^{q_*} p(q,E_1)dq +\int_{q_*}^{q_2} p(q,E_2)dq \right) \right]\eeq
where $q_1$ and $q_2$ can be chosen anywhere in the classically domain of $q$ corresponding
to the states with energy $E_1$ and $E_2$, as $p$ is real there.
The operator ${\cal O}$ does not affect the exponential suppression.
Finally, $q_*$ is the {\em complex} `transition point' that minimises the exponential suppression,
so that $p(q_*,E_1)=p(q_*,E_2)$.

The usual vacuum tunnelling rate is recovered setting $E_1=E_2=0$, so that the two terms in eq.\eq{Landau} merge giving the usual
WKB penetration factor of the classically forbidden potential barrier region.
Eq.\eq{Landau} shows that transitions among different classical states with $E_1\neq E_2$
are generically exponentially suppressed.
This is relevant for vacuum decay induced by particle collisions.

\subsection{Stimulated tunnelling in the thin-wall limit}
Eq.\eq{Landau} was used by Voloshin~\cite{hep-ph/9305219} to compute the stimulated vacuum decay rate
in the thin-wall limit. We here review and validate the computation, and in the next section we try going beyond the thin-wall limit.
A scalar field contains many degrees of freedom.
Its tunnelling reduces to a quantum mechanical problem with one degree of freedom in
thin-wall spherical approximation: the scalar profile is approximated as constant in two regions, 
\beq h(t,r)\simeq \left\{\begin{array}{ll}
h_{\rm true} & \hbox{for $r<R(t)$},\\
h_{\rm false}&\hbox{for $r>R(t)$}.
\end{array}\right.\eeq
The only degree of freedom is  the radius $q(t)=R(t)$ of the thin-wall spherical bubble with surface density $\sigma$
that separates the two vacua with vacuum energy difference $\Delta V = V(h_{\rm false})-V(h_{\rm true})$.
The Lagrangian for $R$ is obtained inserting in the field action a smooth field profile that depends on $(r-R(t))/\delta $,
where $\delta_0$ is the wall thickness at rest, and 
$\delta=\delta_0 \sqrt{1-\dot R^2}$ takes into account its Lorentz contraction.
In the limit of small $\delta_0$, integrating over the volume gives~\cite{Okun}
\beq \label{eq:Lbubble}
L = \int d^3 x\,\Lag =-4\pi R^2 \sigma \sqrt{1-\dot R^2} +\frac{4\pi R^3}{3}\Delta V.\eeq
The  `momentum' conjugated to $R$ is
\beq \label{eq:pR}
p_R = \frac{\partial L}{\partial \dot R} = \frac{4\pi R^2  \sigma\dot R}{\sqrt{1-\dot R^2}} =
\sqrt{\bigg(H+\frac{4\pi R^3}{3}\Delta V\bigg)^2-(4\pi R^2\sigma)^2}\eeq
in terms of the Hamiltonian
\beq \label{eq:Hbubble}
H = p_R\dot R - L= \sqrt{p_R^2 + (4\pi R^2\sigma)^2} -\frac{4\pi R^3}{3}\Delta V.\eeq
A bubble initially at rest, $p_R=0$, expands if $R>R_0 =2\sigma/\Delta V$ corresponding to the critical energy $E_{\rm cr}=16\pi \sigma^3/3\Delta V^2$.
The estimate $r_{0}  \sim 1/\sqrt{\lambda}h_{\rm top}$ of section~\ref{trig} 
is recovered taking into account that $\sigma \circa{>} h_0 \sqrt{\Delta V}$.

\smallskip

Quantising this degree of freedom $R(t)$ allows for quantum tunnelling.
The initial state is the false vacuum with no bubble, corresponding to $R=0$.
The final state is a super-critical bubble.
The tunnelling amplitude is given by eq.\eq{Landau}, that can be conveniently rewritten 
as  \beq \mathscr{A}_{\rm super}=\mathscr{A}_{\rm sub} (E_1,E_2)\mathscr{A}_{\rm tunnel}(E_2)  \eeq
by factoring out $\mathscr{A}_{\rm tunnel}$, the usual WKB amplitude for tunnelling  from the sub-critical to the super-critical bubble.
Then, $\mathscr{A}_{\rm sub} $ is interpreted as 
the amplitude for producing a {\em classically allowed} sub-critical bubble out of the initial state.
\begin{itemize}
\item The tunneling amplitude $| \mathscr{A}_{\rm tunnel}|^2 \approx e^{-S_{\rm tunnel}}$ 
from a thin-wall  sub-critical bubble to a super-critical bubble
is given by the usual WKB computation. 
For vanishing initial-state energy one sets $E_2=0$, and
the classically forbidden region extends from $R_1=0$ to $R_2=3\sigma/\Delta V$, recovering~\cite{Okun}
\beq S_{\rm tunnel}=S_0= 2\int_{R_1}^{R_{2}} |p_R|\, dR =\frac{\pi^2\sigma}{2}R_{2}^3 =\frac{27\pi^2\sigma^4}{2\Delta V^3}\qquad
|p_R|=4\pi R^2 \mu\sqrt{1-\frac{R^2}{R_{2}^2}}.\eeq
For $E_2=E> 0$ the integral gets restricted to the smaller classically forbidden portion of the potential barrier,
so $S_{\rm tunnel}$ decreases reaching zero when $E=E_{\rm cr}$.

\begin{figure}[t]
\begin{center}
$$\includegraphics[width=0.5\textwidth]{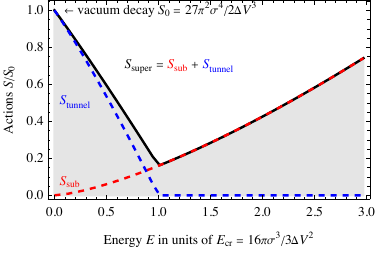}$$
\caption{\label{fig:volo} Exponential suppression factors of vacuum decay induced by particle collisions with energy $E$,
and computed in thin-wall approximation, with surface density $\sigma$ and potential energy difference $\Delta V$.
An exponential suppression remains even at energies large enough that the tunnelling suppression vanishes.}
\end{center}
\end{figure}

\item The amplitude $\mathscr{A}_{\rm sub} $ is obtained inserting in eq.\eq{Landau}
the transition point \beq R_*^3 =-3(E_1+E_2)/8\pi \Delta V\eeq 
with complex $R_*$.
This non-trivial point exists because the Lorentz factor induces a non-trivial Hamiltonian, and it is
determined solely  by the volume term in $p_R$ eq.\eq{pR},
imposing  $(E_1 + 4\pi R^3 \Delta V/3) = - (E_2 + 4\pi R^3 \Delta V/3)$.
Next one sets $E_2=E$ and $E_1=0$, since the energy $E$ initially is in particles, not in the
bubble that approximates the scalar potential energy.
The exponential suppression does not depend on the specific operator ${\cal O}\propto R^3$~\cite{hep-ph/9305219}. 
So $|\mathscr{A}_{\rm sub} |^2 \approx e^{-S_{\rm sub}}$ 
consists of two terms: from $R=0$ up to $R_*$ with energy $E_1=0$,
next from $R_*$ to ${\rm Re}\, R_*$ with energy $E_2$.
\end{itemize}
Fig.\fig{volo} shows our numerical result, that agrees with~\cite{hep-ph/9305219}:
increasing $E$ reduces the exponential suppression $S_{\rm super}=S_{\rm sub}+S_{\rm tunnel}$,
but only partially. The minimum is $S_{\rm super}\approx 0.16 S_0$ at $E=E_{\rm cr}$.
At $E>E_{\rm cr}$  $S_{\rm tunnel}=0$ but the total action $S_{\rm super}=S_{\rm sub}$ grows 
because the computation assumes that all energy must go in
the formation of an unnecessarily big classical bubble.
Vacuum decay stimulated by few-particles collisions remains exponentially suppressed, and thereby negligible.

\subsection{Stimulated tunnelling beyond the thin-wall limit}\label{thick}
The previous section rests on the simplifying thin-wall limit, that affects the result in a crucial way: 
a non-trivial complex transition point exists thanks to the non-trivial relativistic form of the thin-wall action, eq.\eq{Lbubble}.
However the thin-wall approximation only applies when two vacua are nearly degenerate implying a huge exponential suppression.
The approximation does not apply to SM vacuum decay, where the true vacuum can be much deeper than the false SM vacuum.
The literature contains the following two results about stimulated tunnelling beyond the thin-wall approximation:
\begin{itemize}
\item Enqvist computed corrections to the thin-wall approximation at leading order in the thickness of the wall~\cite{hep-ph/9704431},
finding that stimulation with $E>0$  provides a milder reduction (compared to the exact thin-wall limit)
of the exponential suppression of the spontaneous tunnelling rate at $E=0$.

\item Beyond the thin-wall limit, the dynamics does not reduce to a simpler problem with one degree of freedom.
Kuznetsov and Tinyakov~\cite{hep-ph/9703256} employed a coherent-state approximation to
compute thick-wall  vacuum decay induced by particle collisions in a theory with potential $V(h) = m^2 h^2/2 - |\lambda| h^4/4$,
finding that the exponential suppression present at $E=0$ almost entirely persists,
at least up to energies of order $E_{\rm sph}\approx 19m/|\lambda|$.
However the Higgs has a different potential, with sphaleron energy computed in eq.\eq{Esph}.

\end{itemize}
We provide two arguments why particle collisions with energy $E$ negligibly reduce the exponential suppression of  SM Higgs vacuum decay. 

\smallskip

First, we try going beyond the thin-wall limit
by introducing arbitrary reasonable simplifying approximations consisting in assuming special forms for the scalar field profile.
In particular, we assume a profile $h(t,r)=h_0(t)/(1+r^2/R^2)$, quantise $h_0(t)$ and
apply the Landau formalism to complex $h_0$ (rather than complex $R$).
The SM Lagrangian with a constant quartic coupling becomes
\beq \label{eq:Veff}
L(h_0, \dot h_0) = \int d^3 x\, \Lag = \pi^2 R^3 \left[ \frac{\dot h_0^2}{2} -V_{\rm eff}\right]\qquad\hbox{where}\qquad
V_{\rm eff}(h_0)= \frac{h_0^2}{4R^2} +  \frac{ \lambda}{32} h_0^4.
\eeq
The gradient energy produced a barrier term in $V_{\rm eff}$, so that the above approximation can be applied to constant $\lambda<0$.
The  momentum conjugated to $h_0$ is $p_h=\pi^2 R^3 \dot h_0$, giving the Hamiltonian $H = p_h^2/2\pi^2 R^3 +\pi^2 R^3 V_{\rm eff}$.
Tunnelling with $E=0$ gets approximated as
\beq\label{eq:nored}
S_0 = 2 \Im \int_0^{h_0^{\rm Fubini} } p_h\, dh = \sqrt{2} S_{\rm Fubini},\qquad S_{\rm Fubini}=\frac{8\pi^2}{3|\lambda|}
  \eeq
showing that our simplifying assumption is sub-optimal, giving an action $S_0$ larger than the action of the QFT Fubini bounce.
Nevertheless, our simplifying assumption allows to extend the computation to $E>0$ via the Landau eq.\eq{Landau}.
Unlike in the thin-wall limit, the action now has a simple form, so that
the only transition point is $h_* = \infty$ along the real axis, implying that $S_{\rm super}(E) = S_0$.
In other words, the decrease in $S_{\rm tunnel}$ is exactly compensated by the increase in $S_{\rm sub}$,
the suppression needed to convert the particle energy $E$ into a sub-critical bubble.

Our simplified analytic argument gives a result that qualitatively agrees with the numerical computation~\cite{hep-ph/9703256}, 
and is easily extended to the SM potential of eq.\eq{VhSM}.
One obtains the same $V_{\rm eff}$ as in eq.\eq{Veff}
with the quartic $\lambda $ replaced by $ - b \ln(e^{7/6} h_0^2/16h^2_{\rm top})$
i.e.\ renormalized around $h_0$.
Despite this change, the only transition point remains $h_* = \infty$ along the real axis, so $S_{\rm super}(E) = S_0$. 
This result persists assuming the more complicated ansatz $h_0(t)/[1+(r^2-R^2(t))/r_0^2]$.
\nnn{These thick-wall attempts based on simplifying ansatzes cannot capture the full problem,
but suggest an interesting qualitative point: no partial reduction of the exponential suppression;
the reduction found in the thin-wall limit was due to its peculiar dynamics}.

%

\smallskip

We next show that, even if stimulation can mildly reduces the exponential suppression,
in the SM case this would need an energy $E \gg h_{\rm top}$.
The reason is that the SM Higgs potential of eq.\eq{VhSM} is approximatively scale invariant.
As a result, vacuum decay at $E=0$ is well approximated by Fubini instantons 
\beq h(t,r)=\frac{h_0}{1+(r^2-t^2)/r_0^2}\qquad\hbox{with}\qquad
h_0=h_0^{\rm Fubini} = \frac{\sqrt{8/|\lambda|}}{r_0},\eeq 
and generic $r_0$.
The vacuum decay rate is exponentially suppressed by the Fubini
action $S_{\rm Fubini} \approx 8\pi^2/3|\lambda(h_0^{\rm Fubini} )|$ where the quartic coupling is renormalized around $h_0^{\rm Fubini} $.
Since $\lambda(h)$ runs in the SM becoming more negative at $h\gg h_{\rm top}$,
vacuum decay is dominated by field values $h_0^{\rm Fubini}$ much larger than $h_{\rm top}$.
Stimulation by particle collisions with energy $E \sim h_{\rm top}$ can only reduce
the exponential suppression of configurations with $h_0^{\rm Fubini} \sim h_{\rm top}$, 
which are exponentially sub-dominant.
\nnn{Assuming that stimulation with energy $E$ opens a new vacuum decay channel with action 
$S  = \xi \, 8\pi^2/3|\lambda(E)|$,
the overall vacuum decay rate starts being enhanced only above the minimal energy $E\circa{>}  \bp(h_{\rm top}/\bP)^{\sqrt{1-\xi}}$.
The argument around eq.\eq{nored} suggests a $\xi$ around 1 in the Higgs case,
differently from the thin-wall limit where $\xi$ can reach $0.16$.
The reduced Planck mass roughly arises since the SM Higgs quartic runs with $\ln E$ reaching a minimum around $\bP$, 
avoiding a divergence in the instanton size.}

\smallskip

In section~\ref{collider} we clarify an additional possible doubt.

\begin{figure}[!t]
\begin{center}
\includegraphics[width=\textwidth]{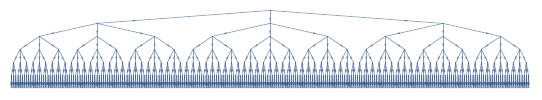}
\caption{\label{fig:multi} Example of a  maximally factorially-enhanced 
Feynman diagram for the production of $N=3^k$ Higgs bosons out of one virtual Higgs using quartic Higgs interactions.
Here $k=6$, giving $N\sim 1000$.}
\end{center}
\end{figure}

\subsection{Vacuum decay via Higgsplosion?}\label{collider}
Studies done before~\cite{Cornwall:1990hh,Goldberg:1990qk,hep-ph/9305219,Voloshin:1992mz,hep-ph/9407381} and after~\cite{1411.2925,1504.05023,1605.06372} 
the Higgs discovery considered the perturbative
cross section $\sigma_N$ for producing $N$ Higgs bosons in collisions with $\sqrt{s}\circa{>} N M_h$, 
finding that it scales as $N! (\lambda/4\pi)^N \sim (N \lambda/4\pi )^N$ for large $N$.
If this behaviour holds up to $N \sim 4\pi /\lambda$ where perturbativity starts failing, the cross section $\sigma_N$ would be large,
a phenomenon dubbed `Higgsplosion'.
If this phenomenon would also hold for ultra-relativistic Higgs bosons,
then a Higgs bubble similar to what needed to trigger vacuum decay would be produced with large cross section.

\smallskip

We here recall the standard `Higgsplosion' argument, extending it to ultra-relativistic Higgs bosons.
For simplicity, we can focus on the Feynman diagram in fig.\fig{multi}
where some unspecified scattering  (for example a $pp$ collision) produced  with amplitude $\mathscr{A}_1$
the first one Higgs boson with virtuality ${s}=E^2$.
For simplicity, we can focus on the symmetric configuration where
the first Higgs splits into 3 Higgses with energy $E/3$ via a quartic interaction.
Iterating $k$ times, one gets $N=3^k$ Higgses with energy $E_1=E/N$.
The diagram in fig.\fig{multi} provides the maximal combinatorial enhancement.
It contains $N_V = (N-1)/2$ quartic vertices
and $N_P =(N-3)/2$ Higgs propagators.
The scattering amplitude is
\beq \mathscr{A}_{N} \sim  \mathscr{A}_{1} \, N!  \lambda^{N_V}  
\Pi
\eeq
where the product of Higgs propagators $\Pi$ does not have a significant $N$ dependence,
being dominated by the most numerous propagators with smaller virtuality $E_1$:
\beq \Pi =\prod_{\ell=1}^{k-1} \frac{1}{[s/3^{2\ell}]^{3^\ell}} = \frac{N^{3}  }{(E_1^2/3)^{N_P}}.\eeq
Taking into account the $N$ identical particles in the final state, the cross section is
\beq
\qquad \sigma_{N}\sim \frac{|\mathscr{A}_{ N}|^2}{s} \frac{\Phi^{(N)}}{N!}
\sim \sigma_1 \left(\frac{N\lambda}{3(4\pi)^2}\right)^{N-1}R_{\rm nr}. \eeq
having written the volume of the $N$-body phase space as $\Phi^{(N)}= \Phi^{(N)}_{\rm rel} R_{\rm nr}$
where
\beq
\Phi^{(N)}_{\rm rel}  = \frac{1}{8\pi} \frac{({E}/{4\pi})^{2N-4}}{(N-2)!(N-1)!}\simeq
E_1^{2N-4}  \frac{(4\pi)^2}{N^2} \left(\frac{e}{4\pi}\right)^{2N},
\qquad
R_{\rm nr}\sim \min (1,\epsilon)^{3N/2}.
\eeq
Here $\Phi^{(N)}_{\rm rel} $ is the ultra-relativistic phase space, and $R_{\rm nr}$ is
the kinematical suppression that arises when $N \sim E/M_h$ is so large that the kinetic energy per particle
$\epsilon = K/M_h =E/NM_h - 1$ is non-relativistic.
\nnn{Our above perturbative results agree with those in the literature, extending them to the values of $E,N$ 
relevant for Higgs vacuum decay, and encountering the same issue:
the cross section gets large only in the regime where perturbativity can no longer be trusted.

\medskip

We thereby proceed discussing arguments from the literature that go beyond the perturbative regime}.

One simple argument considers a similar toy problem in $d=0$ space-time dimensions~\cite{Dine},
where the path-integral that gives the amplitude (up to negligible disconnected diagrams) reduces to the ordinary integral
\beq \label{eq:d0}
\mathscr{A}^{d=0}_N = \frac{1}{\sqrt{2\pi}}
\int_{-\infty}^\infty dh\, h^N\exp\left(-\frac{h^2}{2} - \lambda \frac{h^4}{4}\right).\eeq
Since propagators are $\Pi = 1$ in $d=0$, the `amplitude'  $\mathscr{A}^{d=0}_N $ counts the number of diagrams.
Its perturbative expansion gives the same $N!$ enhancement as in $d=3+1$, and breaks down at $\lambda N\circa{>} 1$.
The non-perturbative eq.\eq{d0} shows that the toy amplitude $\mathscr{A}^{d=0}_N $ remains exponentially suppressed:
it can be approximated by moving $h^N$ into the exponential, finding that the `saddle point' that minimises the
effective action has $h\approx (N/\lambda)^{1/4} $, away from the perturbative expansion point $h=0$.
This argument clarifies the physics in $d=3+1$, 
suggesting what happens \nnn{in the deep non-perturbative regime}:
a large number $ N\gg 4\pi/\lambda$ of Higgs quanta
forms a classical configuration that invalidates 
the perturbative expansion around the vacuum (see also~\cite{1909.01269}).
However, this first argument does not tell if, in the physical Higgs problem,
the cross section $\sigma_N$  gets large  at $ N \sim 4\pi /\lambda$, \nnn{around the boundary of the perturbative regime}.
This is the relevant regime for stimulated vacuum decay.

The first computation that covers this critical regime in $d=3+1$ was performed in~\cite{2212.03268} 
using a method developed in~\cite{Rubakov:1991fb,Rubakov:1992ec,hep-ph/9505338}, finding 
an exponentially suppressed cross section.

\medskip

\nnn{In conclusion, the cross section for producing configurations with $N$ Higgs bosons relevant for Higgs vacuum decay
is exponentially suppressed as naively expected.
This is not a trivial result, as in other cases these exponential suppressions are compensated
by the multiplicity of the macroscopic state.
For example, black hole production from super-Planckian collisions is not exponentially suppressed 
(see e.g.~\cite{hep-ph/0106295,hep-ph/0106219} for computations in models where large extra dimensions allow black hole production at lower energy); a special  `classicalization' mechanism could take place in the Higgs case
adding ad-hoc new physics~\cite{1010.1415}.
We stick to the Standard Model.}

\section{Highest-energy cosmic-ray collisions}\label{CR}
In section~\ref{bubble} we found that vacuum decay stimulated by few-particle collisions remains exponentially suppressed
by a factor of order $S_{\rm Fubini}\sim 1000$.
We here estimate that the number of cosmic-ray collisions with energy $\sqrt{s}$ comparable to the SM vacuum instability scale
is below $e^{160}$ even under most optimistic assumptions.
So we will conclude that cosmic rays would not have triggered SM vacuum decay.

\smallskip

{\sc Auger} (and the smaller TA experiment in the northern hemisphere) 
detected cosmic rays up to energy $E\approx 2~10^{11}\GeV$ with flux roughly given by~\cite{1807.09645,1903.06714,2212.01600}
\beq\label{eq:CR} \frac{d\Phi}{d\ln E} \approx \frac{{\rm EeV}^2}{E^2} \frac{100}{{\rm km^2}\,{\rm yr}\,{\rm sr}}\eeq
but could not clarify their composition,
finding something intermediate between proton and nuclei.
Furthermore, they could not identify the production sites of ultra-high-energy CR  (hints are claimed in~\cite{1801.06160,2007.00023}).
Our numerical computations will use a precise CR flux, raher than eq.\eq{CR}.

\subsection{Collisions of 2 cosmic rays}

The CR collision with highest energy ever occurred has been estimated to be $\sqrt{s} \sim 10^{11}\GeV$~\cite{Hut:1983xa}.
Indeed, the CR number density $n(E)= 4\pi \int_E^\infty dE \, d\Phi/dE$ implies that the number of CR collisions
with energy $s\circa{>} E^2$ is
\beq N_2(s) \sim T_{\rm U} R^3_{\rm U} \sigma_2 n^2 \sim (E/{\rm EeV})^6\eeq
having integrated over the universe time and volume $T_{\rm U}\sim R_{\rm U} \sim 10\,{\rm Gyr}$
and adopted $\sigma_2(s)=1/s$ as a reference cross section among two CR~\cite{Hut:1983xa}
(this is roughly appropriate if CR have a non-negligible proton component).
This estimate was employed by collider safety reports such as~\cite{hep-ph/9910333,0805.4528,0806.3414}.
A more precise expression is\myfootnote{We collide $P_1 = E_1(1, 1,0,0)$ with $P_2 = E_2 (1, - \cos\theta, -\sin\theta,0)$.
So $s=(P_1+P_2)^2 =2 E_1 E_2 (1+c)$.
The Lorentz-invariant $v_{\rm rel}$ equals 1 in the ultra-relativistic limit,
while the event rate is proportional to 
$ \sqrt{(\vec{v}_1-\vec{v}_2)^2 - (\vec{v}_1\times\vec{v}_2)^2} = 1-\cos\theta$.
The event rate is
\beq N_{\rm ev} = \int d^4x  \int_{-1}^{+1} \frac{dc}{2}\, (1-c) dE_1 \,dE_2 \sigma(s) \frac{dn}{dE_1}\frac{dn}{dE_2}.\eeq
We switch integration variables to $s$ and $r\equiv E_1/E_2$ obtaining
\beq \frac{dN}{ds} = \int d^4 x \int_{-1}^{+1} \frac{dc}{2} (1-c)\int \frac{dr}{2ry} 
\frac{dn}{dE}\left(\sqrt{\frac{s}{ry}}\right)
\frac{dn}{dE}\left(\sqrt{\frac{rs}{y}}\right) \sigma(s)\eeq
where $y=\sqrt{2(1+c)}$.
If $db/dE \propto 1/E^\gamma$ the angular dependence separates leaving  $dN/ds \propto \sigma\, s^{-\gamma}\int dr/r$.
The log-divergent $\int dr/r$ is cut adding an exponential-like cut-off to the spectrum.
This can be better written as
\beq \frac{dN}{d\ln s}= \int d^4 x \int_{-1}^{+1} \frac{dc}{2} (1-c)\int d\ell 
\frac{dn}{d\ln E}\left(\frac12 \ln \frac{s}{2(1+c)} -\frac{\ell}{2} \right)
\frac{dn}{d\ln E}\left(\frac12 \ln \frac{s}{2(1+c)} +\frac{\ell}{2} \right)
\sigma(s)
\eeq
where $\ell=\ln (E_1/E_2)$.}
\beq \frac{dN_2}{ds} = B_2 \int d^4 x \int_{-1}^{+1} \frac{dc}{2} (1-c)\int \frac{dr}{2ry} 
\frac{dn}{dE}\left(\sqrt{\frac{s}{ry}}\right)
\frac{dn}{dE}\left(\sqrt{\frac{rs}{y}}\right) \sigma(s)\eeq
where $\int d^4 x \approx 16\pi/1485 H_0^4$ is the volume of our past light-cone (neglecting the late-time acceleration),
$r=E_1/E_2$ is the ratio between the collision energies $E_1$ and $E_2$, $c=\cos\theta$ is the collision angle, and $y=\sqrt{2(1+c)}$.
Apart from  averaging over the angle, we introduced a `boost factor' $B_2$ that accounts for  CR inhomogeneities.
Since CR are likely produced in astrophysical sources, this enhances the CR collision rate in a significant way.
Let us assume that CR 
have a higher density around $N_{\rm s}$
production sources with size $R_{\rm s} \ll R_{\rm U}$ and duration $T_{\rm s}$,
where CR stay for a time $\tau_{\rm s}$.
Then the CR density around one source is $n_{\rm s}\sim  n(R_{\rm U}/R_{\rm s})^3 (T_{\rm U}/T_{\rm s})/N_{\rm s}$, and the boost factor is
\beq B _2 \approx \frac{\med{n^2}}{\med{n}^2} \sim   \frac{1}{N_{\rm s}} \frac{\tau_{\rm s} T_{\rm U}}{T_{\rm s}^2}\frac{ R^3_{\rm U}}{R_{\rm s}^3 } \eeq 
where $\med{}$ denotes the average over space and time.
The values of $R_{\rm s}$, $T_{\rm s}$, $\tau_{\rm s}$ are unknown, as
{\sc Auger} and TA could not identify the production sites of ultra-high-energy CR (see~\cite{1801.06160,2007.00023} for hints),
and the plausible possibilities significantly differ.
In particular, smaller $R_{\rm s}$ increases $B_2$.
If multiple CR sources contribute comparably, the largest enhancement applies to the boost factor.
Plausible production sites are~\cite{1807.09645,1903.06714}:
\begin{itemize}
\item The smallest sources are pulsars of magnetar type, 
neutron stars rotating with angular velocity $\omega_{\rm s}$
formed from supernov\ae{} collapses that have
the largest magnetic fields $B_{\rm s}\circa{<} 10^{11}\,{\rm T}$,
thereby allowing acceleration up to $ E \circa{<} e B_{\rm s} R_{\rm s} (\omega_{\rm s} R_{\rm s})^2$~\cite{1801.06160} 
in a small acceleration region with size
comparable to the Schwarzschild radius, $R_{\rm s} \sim 2 G M \sim 10\,{\rm km}$ for $M \sim ( 1.4-10) M_{\rm sun}$.
About $N_{\rm s}\sim 10^{19}$ such objects are estimated to exist in our past light-cone, and
their magnetic field decays  in $T_{\rm s}\sim 10^4\,{\rm yr}$.
Assuming the minimal $\tau_{\rm s}\sim R_{\rm s}$,
this leads to $B_2 \sim 10^{38}$, or even to $B_2 \sim 10^{59}$ if a significant part of CR acceleration happens just after the collapse,
in $T_{\rm s}\sim 10\,{\rm s}$.
Loosely similar estimates apply if CR are accelerated during $\gamma$-ray
bursts from various kinds of supernov\ae~\cite{2007.06409}.

\begin{figure}[t]
\begin{center}
\includegraphics[width=0.74\textwidth]{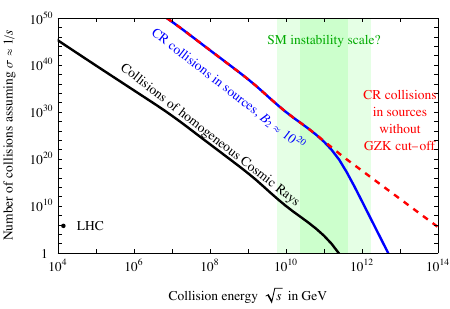}
\caption{\label{fig:NumberCRcollisions}  Number of collisions of two cosmic-rays:
a) in the minimal uniform assumption (black curve);
b) including a boost due to collisions in small CR acceleration sites, $B_2 \sim 10^{20}$ (blue curve,
$B_2$ up to $10^{59}$ are discussed in the text);
c) assuming no GZK cut-off in CR acceleration sites.}
\end{center}
\end{figure}

\item The larger sources are Active Galactic Nuclei~\cite{1404.3176}, 
objects with rotating magnetospheres around super-massive galactic black holes of mass $M \sim 10^8 M_{\rm sun}$ and
duration $T_{\rm s} \sim 10^5$ yr~\cite{1505.06733}.
Given that our horizon contains about $10^{12}$ galaxies,
we estimate  $N_{\rm s}\sim 10^{11}$ super-massive galactic black holes.
This gives $B_2 \sim 10^{28}$ if acceleration happens in the inner region around the
Schwarzschild radius $R_{\rm s} \sim 2 GM \sim 10^3\,{\rm s}$,
or $B_2 \sim 10^6$ if acceleration happens in a galactic-size Mpc region.
Much larger $B_2 \sim 10^{35}$ could arise if CR acceleration significantly happens in brief events,
while the black hole accretes nearby stars by tidal disruption~\cite{1903.06714}.

\end{itemize}
These objects tend to have a  geometry favourable to high-energy collisions,
as incoming accelerated particles are channelled towards out-going jets at magnetic poles.

Assuming a negligible boost, $B_2\sim 1$,
fig.\fig{NumberCRcollisions} shows that the CR maximal collision energy is around the
cut-off observed by {\sc Auger} and TA, and possibly due to the 
GZK effect (the $p\gamma_{\rm CMB}\to \Delta^+$ process opens at high energy and
leads to absorption
on distances larger than $\sim 100\,{\rm Mpc}$).
This collision energy is comparable to the uncertain SM Higgs instability scale.
Fig.\fig{NumberCRcollisions} shows that a `modest' boost factor $B_2 \sim 10^{20}$ brings the maximal
collision energy above the Higgs instability scale.

Finally, CR around production sources could reach higher energies not limited by the GZK effect.
In such a case, significantly larger collision energies happened (red curve in fig.\fig{NumberCRcollisions})
if around production sites the CR energy spectrum is a power law with no GZK cut-off.
Some other cut-off is however expected from the acceleration mechanism itself, 
and maybe this (rather than the GZK effect) generated the cut-off observed in CR.


\subsection{Collisions of $N$ cosmic rays}
Collisions among $N\gg 1$ initial-state particles are relevant for vacuum decay.

Starting from $N=3$, the number of 3-collisions among cosmic rays with energy $E$ is $N_3 = B_3\int d^4x\, n^3\sigma_3$ 
where we can estimate the 3-particle `cross section' as $\sigma_3 \sim 1/E^5$ and the boost factor as
\beq  B _3 \approx  \frac{\med{n^3}}{\med{n}^3} \sim  B_2 B_1\qquad B_1 =  \frac{n_{\rm s}}{n} \sim  \frac{T_{\rm U} R_{\rm U}^3}{N_{\rm s} T_{\rm s} R_{\rm s}^3}.\eeq
The plausible CR accelerations sites discussed in the previous section lead to the following extreme cases.
At one extremum, CR accelerated by AGN on Mpc scales have a mild $B_1\sim 10^5$ resulting in a negligible $N_3\ll 1$ at GZK energies
comparable to the SM instability scale.
At the opposite extremum, CR accelerated by small magnetars on short time-scales 
can have $B_1\sim 10^{60}$, implying $N_3  \gg 1$ at GZK energies.

\smallskip

At larger $N$, the boost factors grow as $B_N/B_{N-1} \sim  B_1$ resulting in the number of $N$-collisions
$N_N /N_{N-1} \sim B_1 n/E^3  \sim (E_0/E)^5$.
The energy $E_0$ is low, $E_0 \sim 10\GeV$, even in the most optimistic case of small magnetars.
This means that collisions of a large number $N \sim 1000$ of cosmic rays never occurred at energies comparable
to the SM instability scale.



\section{Vacuum decay triggered by many particles}\label{many}
We have seen that, to avoid exponentially suppressed rates,
Higgs vacuum decay must be triggered by an initial state that contains a large number $N\circa{>} 10^3$ of Higgs quanta.
We conclude with futuristic and unrealistic speculations 
about the possibility of artificial large $N$ collisions at ultra-high energy.
Despite being at hypothetical level, the subsequent discussion perhaps has theoretical interest, as it will involve some basic fundamental issues.

\subsection{Right-handed electron beams}
As a first attempt, we consider the possibility of triggering Higgs vacuum decay by  colliding into one center
beams made of the particles or anti-particles so far used by colliders: protons or electrons (with muons being a future option).
However, this collision risks producing all SM particles (not only the Higgs), 
approximatively giving a thermal-like environment that `burns' the Higgs instability,
as the potential $V(h)$ would be contain an extra stabilising 
thermal-like Higgs mass term $M_h^{\rm th}=\kappa T$, similarly to what happens at finite temperature~\cite{Arnold:1991cv}.
Here $\kappa^2 \approx (g_Y^2+3 g_2^2)/16 + y_t^2/4 + \cdots \approx 0.35^2$ is a combination of SM couplings~\cite{Arnold:1991cv}. 
As SM couplings are perturbative, collisions between many particles occur with negligible rate in the
finite temperature plasma that filled the early universe.
The thermal system can be computed via Euclidean methods, finding 
that particle collisions lead to 
a vacuum decay rate exponential suppressed by a factor $S_{\rm thermal} \approx 6.015\pi \kappa /|\lambda|$~\cite{Arnold:1991cv}.

\smallskip

Colliding beams made of particles only would initially provide a chemical potential that tends to destabilise the Higgs vacuum.
However, interactions among ultra-relativistic particles risk generating dominant thermal effects.
To mitigate the unwanted effects, we consider beams of ultra-relativistic electrons with right-handed polarization,
so that only hyper-charged fields are sourced,
and the electron Yukawa coupling is small enough that the $e_R$ charge is approximately conserved.
A steady stream of $\dot N$ right-handed electrons per unit time that enter and leave a central zone at $r\approx 0$ will be necessary.
So we approximate the system as spherical and quasi-static with density
$n_e = \dot N/2\pi (r^2 + r_c^2)$ up to some outer radius $R_{\rm out}$.
The core radius $r_c$ arises if electrons reach the focus with impact parameters up to $r_c$.
The maximal $e_R$ arrival rate into a sphere with radius $r_c$ allowed by Pauli blocking is
\begin{equation}\label{eq:dotNe}
\dot N\lesssim4\pi r_c^2\int_{\rm inward}\frac{d^3p}{(2\pi)^3}\,\cos\theta
=\frac{(p_c r_c)^2\,\Delta p}{2\pi}\sim\frac{p_c^3r_c^2}{2\pi}.
\end{equation}
So electrons need to have ultra-relativistic momentum $p_c$ at the focus and a broad spectrum $\Delta p\sim p_c$,
together with angular dispersion.
The hyper-electric field is given by the usual Gauss result, since
relativistic effects, magnetic fields and radiation cancel out due to spherical symmetry, leaving
$E_{Br}(r)=-g_Y N_{\rm enc}(r)/4\pi r^2 = - \partial\varphi_B/\partial r$, where
$N_{\rm enc}(r) = \int_{r'<r} d^3r' \,n_e$ is the number of $e_R$ inside radius $r$.
The dimensionless strength of this hyper-electric field is $Z\equiv g_Y^2\dot N r_c/8\pi$,
analogous to $Z\alpha$ for a nucleus.
In practice, $N_b=4$ beams with $r_c$ size along tetrahedral directions
can be approximated as spherical up to $1/N_b$ corrections.


The hyper-electric field triggers $H \bar H$ pair creation of the quanta of the Higgs doublet $H$.
Higgs particles feel a potential $U_H (r,t)= (g_Y \varphi_B - g_2 \varphi_{W_3})/2$.
For later convenience we included the electric potential of the weak $W_3$ boson,
and work in the gauge where the vector potentials $A_r$ vanish.
So $H$ are attracted, while $\bar H$ are repelled and escape.
Then $H$ forms a condensate: this is the Schiff-Snyder-Weinberg phenomenon and the supercritical Bose condensation
of Klein and Rafelski \cite{SSW,KR75}, as described by the Klein-Gordon equation
\beq \label{eq:RotatingH}
\left(\frac{\partial}{\partial t}+iU_H\right)^2 H -\nabla^2H+\Big[\frac{V'(h)}{h}+M_h^2\Big]H=0 \eeq
where $H = (0,h e^{i\theta}/\sqrt{2})$, $h$ is the radial mode,
$V(h)$ is its SM potential, and
$M_h^2$ is the Higgs mass, made significant by medium effects discussed below.
The key features of eq.\eq{RotatingH} are that the variation of $U_H$ across the well creates $H\bar H$ pairs,
and positive-energy $H$ states inside the well overlap with negative-energy $\bar H$ states outside.
Unlike Schwinger pair production in a uniform field, 
where the created particles fly apart, here the $H$ quanta are trapped in the well,
and this makes the computation simple. 
The quantum vacuum only supplies the initial seed (half a quantum in the growing mode);
each new $H$ then falls into the same bound mode and stimulates further emission into it, as in a laser.
So the Higgs occupation grows exponentially as $e^{2\gamma t}$ rather than linearly in time,
and becomes large enough to be described by the classical field equation.

Initially, when the condensate is small and $U_H(r)$ is the static potential of the source, the growth rate of the
Higgs condensate is $\gamma={\rm Im}\,\omega$, where $\omega$ is found by solving eq.\eq{RotatingH} for an $s$-wave
$H\propto e^{-i\omega t}u(r)/r$,
with outgoing boundary conditions which for ${\rm Im}\,\omega>0$ decay at large $r$:
\beq\label{eq:Hcondensate}
\left[\frac{d^2}{dr^2}+(\omega-U_H(r))^2-M_h^2\right]u=0,\qquad u(0)=0,\qquad u(r\to\infty)\propto e^{i\omega r}.
\eeq
However, the hyper-electric field also creates SM fermion pairs that screen the $U_H$ potential.
This unwanted effect is Pauli limited.
To estimate the largest possible screening, we assume that all bound fermion states are filled up to zero energy.
We describe this filling in the relativistic Thomas--Fermi approximation,
treating the 45 Weyl fermions of the SM as massless and at zero temperature.
Before the condensate forms ($\varphi_{W_3}=0$), each fills momenta $p<|2Y_f U_H|$,
so the Poisson equation for $U_H$ gets one extra negative term proportional to $U_H^3$:
\begin{equation}
\nabla^2U_H=\frac{2Z}{r_c\,(r_c^2+r^2)}\,\theta(R_{\rm out}-r)+\frac{g_Y^2}{2}\,\frac{8\sum_f|Y_f|^4}{6\pi^2}\,U_H^3 .
\end{equation}
Ignoring fermion screening and the Higgs mass,
eq.\eq{Hcondensate} gives $\gamma \approx 0.08 Z^{1/3}/r_c$ for $Z\gtrsim 3$,
roughly independent of $R_{\rm out} \gtrsim r_c$.
The fermionic screening, which builds up faster than the condensate, reduces $\gamma$ by a factor 1.2--1.8.
Furthermore, the medium density of top quarks, $t_L$ ($Y=\frac16$, $T_3=\frac12$) and $t_R$ ($Y=\frac23$)
induces a term $y_t h \med{t \bar{t}}/\sqrt{2}$ which results in a positive Higgs mass
\begin{equation}\label{eq:topinducedMh}
M_h^2 \approx \sum_{t_L,t_R}3\,\frac{y_t^2}{2}\,\frac1{2\pi^2}\int_0^{p_F}\frac{p^2\,dp}{\sqrt{p^2+m_t^2}},\qquad
p_F=\sqrt{V_t^2-m_t^2},
\end{equation}
where $V_t=g_YY_t\varphi_B+g_2T_3\varphi_{W_3}$ is the potential energy of the top and $m_t=y_th/\sqrt2$
(no tops are trapped where $m_t>|V_t|$).
Top mixing, omitted in eq.\eq{topinducedMh}, is included in the numerical computation.
This reduces to $M_h^2\approx0.07\,y_t^2U_H^2$ neglecting the top mass.
For the measured value of the top quark Yukawa coupling this $M_h^2$ still allows a rate $\gamma>0$.

So a sustained $e_R$ stream triggers the formation of a Higgs condensate exponentially growing with rate $\gamma$.
As discussed previously, the SM Higgs instability is triggered if $h \gtrsim h_{\rm top}$ and
if the number of Higgs quanta is $N_H \gtrsim 1500 \approx e^{7.3}$.
Starting from the vacuum seed (half a quantum), this requires the growth $N_H \propto e^{2\gamma t}$
to persist for $N_{\rm ef} \approx4$ $e$-folds, or somewhat more if the growing mode extends beyond the critical radius.
When $h$ reaches $\approx 0.2\, h_{\rm top}$ the Higgs condensate is so intense that it starts screening its source;
intensifying the source needs higher-energy $e_R$ in view of eq.\eq{dotNe}.
We neglected scatterings, heating, possible other electroweak condensates; Higgs decays are largely Pauli blocked.

\begin{figure}[t]
\begin{center}
$$\includegraphics[width=0.45\textwidth]{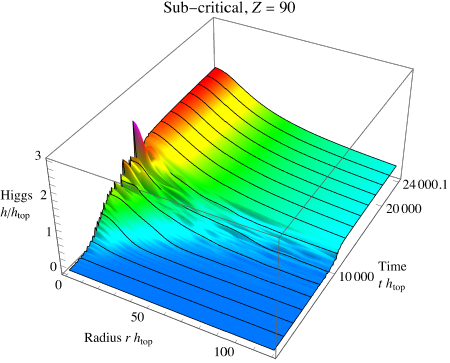}\qquad\includegraphics[width=0.45\textwidth]{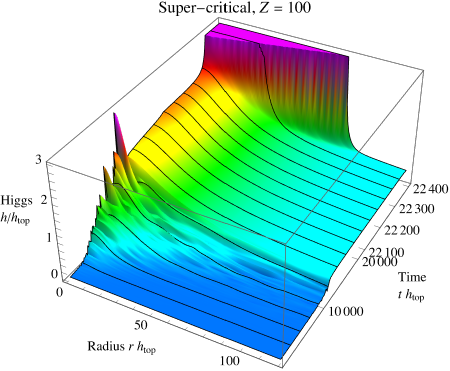}$$
\caption{\label{fig:banghikeR} Classical evolution of a spherical Higgs field $h(t,r)$ triggered by a
continuous spherical stream of right-handed electrons with strength $Z$, for $y_t=0.5$ and $r_c=16/h_{\rm top}$.
a) For $Z=90$ the Higgs condensate grows and saturates without triggering vacuum decay;
b) For $Z=100$ the condensate nucleates a super-critical bubble that ignites vacuum decay,
as emphasized by the non-linear time scale.
The oscillations are physical breathing modes of the condensate, not numerical noise.
}
\end{center}
\end{figure}

At this point multiple effects take place and order-unity factors are relevant.
We thereby solved eq.\eq{RotatingH} numerically, assuming
that Thomas–Fermi adjustment is so fast that we can neglect the associated currents.
We used $r_c=16/h_{\rm top}$, $R_{\rm out}=10\,r_c$ and $y_t=0.5$, the SM top Yukawa coupling renormalized around the Higgs instability scale.
Fig.\fig{banghikeR} shows the result: within the approximations, triggering vacuum decay is possible for large enough $Z$.
The condensate first saturates at $h\approx 2\,h_{\rm top}$ in a breathing configuration.
For $Z=100$ it nucleates a true-vacuum bubble after $t\approx 2~10^4/h_{\rm top}$, which then expands at the speed of light,
while for $Z\lesssim 90$ no decay occurs.
$Z=100$ corresponds to $\dot N\approx 1.6~10^4/r_c$, so Pauli blocking requires
$p_c\approx 3\,h_{\rm top}$ and, adding the hyper-electric potential barrier, an injection momentum $\approx 5\,h_{\rm top}$.
Sustaining the stream until vacuum decays requires $\approx 2~10^7$ electrons, for a total energy
$\approx 10^8\,h_{\rm top}\approx 3~10^5 E_{\rm sph}$, most of which is carried away by the outgoing electrons.
The probability that an electron scatters during its pass is small,
$\wp \approx \sigma_{e_R e_R}\dot{N}/4 r_c\sim \hbox{few}\, \%$.

We made approximations, chose favourable source parameters, and found that order-unity factors are critical.
A full computation is needed to establish whether an optimised configuration can trigger vacuum decay.
An order-unity factor can be gained by adding converging beams of right-handed muons.


%

\subsection{$N$ Higgs beams}
The above discussion indicates that triggering vacuum decay might need a cleaner initial state of $N$ Higgs quanta,
not accompanied by a much larger number of SM particles, as they would effectively generate a Higgs thermal mass.
We wildly speculate if, in line of principle, these $N$ Higgs
could be produced  by $N$ muon colliders `boosted'  up to unrealistic energy.
One lepton collider with energies $E_{\ell^\pm}$ and collision angle $\theta \ll 1$
can produce resonantly 
one boosted Higgs with mass $ M_h =  \sqrt{E_{\ell^-} E_{\ell^+}}\theta$ and energy $ E_h =E_{\ell^-}+ E_{\ell^+}$.
This collision geometry and its possibile luminosity was studied in~\cite{2306.12480} at realistic TeV energies,
irrelevant for Higgs vacuum decay.
Using $N$ such colliders with unrealistically higher energy $E_{\ell^\pm}\sim 10^{9}\GeV$,
one could engineer $N$  Higgs bosons simultaneously hitting a central interaction point,
while SM particles produced by other processes with comparable cross sections $\sigma \sim g^4/4\pi M_h^2$
are not focused into this center by the resonant kinematics.
The distance from the production point to the interaction point must be smaller than the life-time of the boosted Higgs bosons, 
\beq \tau_h  \frac{E_h}{M_h}= 0.37 \,{\rm \mu m}\,\frac{E_h}{10^{9}\GeV}.\eeq
Achieving the instantaneous luminosity needed to trigger vacuum decay
risks conflicting with a fundamental issue.
Existing machines collide $N=2$ beams made of bunches of particles with transverse size $\Delta x \sim \mu{\rm m}$
because each one more beam suppresses the $N$-event rate by a macro/micro factor $s \,\Delta x^2$, where $\sqrt{s}$ is the collision energy.
The luminosity ${\cal L}$ achieved by current technology is about 15 orders of magnitude than the quantum limit:
the maximal luminosity achievable by colliding two fermionic particles is
\beq \label{eq:Lumith}{\cal L} \sim \frac{1}{\Delta t \,\Delta x^2} \circa{<} \Delta E \,\Delta p^2\eeq
in view of the uncertainty principle $\Delta E\, \Delta t \circa{>}1 $ and $\Delta x\,\Delta p \circa{>}1$.
This shows how two well-known features of realistic collisions among two bunches of particles follow from fundamental physics:
luminosity growing as $s$, and linear trade-off between luminosity and beam energy spread $\Delta E$.
As a result of eq.\eq{Lumith}, 
a generic two-particle crossing produces one collision event with probability $\sim g^2$ in a SM-like theory with dimension-less
couplings $g$.
The above estimate is practically realized at finite temperature $T$:
according to Fermi-Dirac statistics, fermions have order unity occupation numbers at $E\circa{<} T$. 
SM thermal field theory is computable perturbatively, since $g \circa{<} 1$ suppresses $N$-particle collisions.

\smallskip

We here focus on a specific favourable collision: resonant on-shell production, such that cross sections
can reach the unitarity limit,  $\sigma_{\rm peak}\sim 4\pi/s$, corresponding to non-perturbative couplings.
However, two suppression factors arise in the case at hand.
First, the $\mu^-\mu^+\to h$ cross section is 
$\sigma_{\rm peak} = 4\pi R\, {\rm BR}(h\to \mu^+\mu^-)/M_h^2$,
where $R\approx0.5$ accounts for initial-state radiation and ${\rm BR}(h\to \ell^+\ell^-)\approx 0.22~10^{-3}$ in the SM.
Second, achieving the Higgs resonance needs $\Delta E \circa{<}\Gamma_h$ where $\Gamma_h$ is the Higgs decay width.
With the proposed geometry, the maximal number of events per one $\mu^-\mu^+$ collision is
$ \sim \Gamma(h\to \mu^-\mu^+)/M_h$, implying that
colliding $N$ Higgs bosons needs $\gg N$ colliders.

In line of principle, this quantum limit on the luminosity
would be avoided by a boson collider, for example able of colliding overlapped photon quanta as $\gamma\gamma\to h$.
Needless to say, this is even more unrealistic: laser beams have  been produced at atomic eV energies, not at $10^{9}\GeV$.

The possibile utility of this speculative discussion is illustrating how, even if the Higgs vacuum is unstable,
a significant safety factor is built in fundamental physics.
Similar considerations limit the possibility of producing
solitonic configurations at the weak scale~\cite{1910.04761}, as well as the possibility of triggering
other unknown vacuum decay channels that might hypothetically exist at current energies.

\begin{figure}[t]
\begin{center}
$$\includegraphics[width=0.45\textwidth]{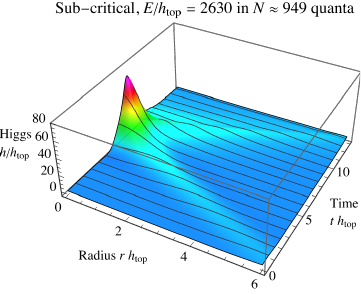}\qquad\includegraphics[width=0.45\textwidth]{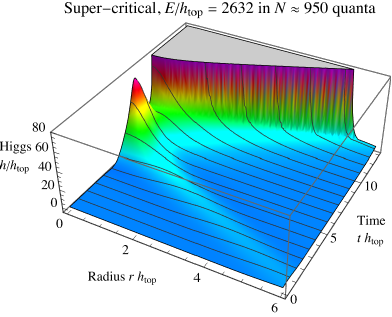}$$
\caption{\label{fig:banghik} Classical evolution of an in-coming spherical Higgs wave containing  about 1000 ultra-relativistic Higgs quanta 
with individual energy $k$ peaked around  $h_{\rm top}$ and total energy $E$. 
a) Forming a slightly sub-critical bubble that dissolves, despite reaching $h \gg h_{\rm top}$;
b) Forming a slightly super-critical bubble that ignites vacuum decay.}
\end{center}
\end{figure}

\subsection{Classical Higgs wave}
We proceed with our discussion showing how
triggering vacuum decay would be possible if a beam of $N\gg 1$ Higgs bosons were hypothetically available.
Such a multi-particle system could be approximated as a (theoretically simple) classical inward-going spherical Higgs wave.
Triggering vacuum decay via classical evolution would avoid the exponential suppression of the quantum rate.

We recall the classical Higgs equation in flat space,
$\ddot h - h'' - 2h'/r = -V'$
for a spherical wave $h(t,r)$. 
We denote as $r=r_0$ the radius at the points 
where the Higgs bosons are produced at $t=0$.
In terms of $u(t,r)=r\, h(t,r)/r_0$ the wave equation becomes $\ddot u - u'' = - rV'/r_0$.
Around the production surface $r\approx r_0$ we can initially neglect the Higgs potential $V(h)$, 
because the Higgs field value is well below $h_{\rm top}$ and because the Higgs quanta are ultra-relativistic. 
So the time evolution of the inward wave is initially approximatively solved as \beq h(t,r) \simeq u_0(r+t)  r_0/r.\eeq
Assuming as initial profile $u_0$ a short wave-packet with length $1/k$ peaked at $r\approx r_0$,
the energy $E =\int 4\pi r^2 \, dr [\dot h^2/2 + h'^2/2 + V]$ of the full system
is roughly given by $E \sim 4\pi r_0^2 k h_0^2$, 
corresponding to $N \approx E/k = 4\pi r_0^2 h_0^2$ Higgs quanta.
Here $h_0 = h(0,r_0)$, where $t=0$ is the $\ell^-\ell^+\to h$ production time.

\begin{figure}[t]
\begin{center}
$$\includegraphics[width=0.45\textwidth]{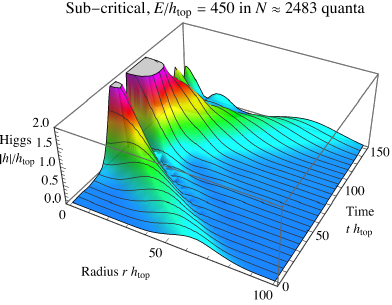}\qquad\includegraphics[width=0.45\textwidth]{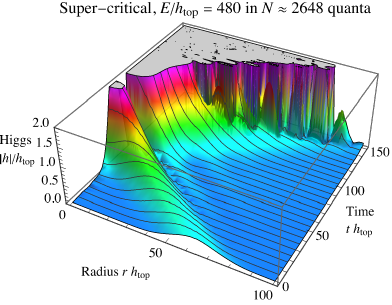}$$
\caption{\label{fig:banglowk} As in fig.\fig{banghik}, but using about 2500 Higgs quanta with lower energy $k $ peaked around  $0.01 h_{\rm top}$,
so that a bubble with larger radius $\sim 1/k$ and lower field value $h \approx h_{\rm top}$  ignites vacuum decay. }
\end{center}
\end{figure}

In fig.\fig{banghik} and\fig{banglowk} we show the numerical solution to the classical evolution equations, choosing an incoming wave-packet
with arbitrary profile of the form
\beq u_0 (r)=h_0 \sin[k(r-r_0)]e^{-k (r-r_0)^2/2}\eeq
dependent on two free parameters: $k$ (that controls the typical energy of one Higgs quantum), and $h_0$ (that controls the intensity of the wave).

Fig.\fig{banghik} shows the evolution using quanta with $k = h_{\rm top}$ i.e.\ energies around $h_{\rm top}$.
In the left panel we see how a spherical in-ward wave forms a bubble with $h> h_{\rm top}$
that however does not trigger vacuum decay and dissolves into an out-ward wave.
Increasing the intensity of the wave crosses the critical value above which the bubble ignites vacuum decay.
This case is shown in fig.\fig{banghik}b, and corresponds to roughly the minimal number $N \sim 1000$ of Higgs quanta.

\smallskip

Fig.\fig{banglowk} shows the similar result using quanta with lower energy $k$ peaked around $0.01 h_{\rm top}$.
This only allows to form bubbles with large radius $r_0\sim 1/k \sim 100 h_{\rm top}$, that ignite vacuum decay 
as soon as the Higgs field
inside is slightly above the top of the SM potential barrier at $h_{\rm top}$.
Compared to the previous case of fig.\fig{banghik}, the total energy of the configuration is lower, $E\sim 500 h_{\rm top}$,
but a larger number of Higgs quanta $N \sim 2600$ is needed.
While in principle colliders with $k \sim \TeV$ can trigger vacuum decay, 
even this possibility is unrealistic, as it would need a huge number $N \sim (h_{\rm top}/k)^3$
of Higgs bosons that would decay too fast, with low boost factor.

\smallskip

To avoid hitting a singularity at $h=\infty$, our numerical computations introduced an extra
non-renormalizable term in the Higgs potential that creates a deep minimum at large field value.
If this is absent and/or deep, gravity becomes relevant as the Higgs falls towards the true vacuum.
As shown in~\cite{1607.00381,2207.00299}, gravity cannot stop the explosion:
a black hole can form inside the bubble, that anyhow expands at the speed of light.

\subsection{Can a vacuum bubble be controlled, extracting energy?}\label{energy}
To conclude we bring the discussion to an even more speculative level,
that could nevertheless be intellectually interesting because a new issue of basic fundamental physics arises:
if a vacuum instability exists and can be triggered, 
{\em is it possible to use vacuum energy as an energy source, controlling the energy release}?
The answer is naively negative because vacuum energy, unlike energy stored in atoms or nuclei, is not limited by the amount of matter:
the release of vacuum energy could be unstoppable.


We next discuss how, at least in theory, the answer could be positive.
In line of principle, the expansion of a deeper Higgs vacuum bubble could be slowered or blocked by 
hitting all its surface with beams of SM particles intense enough to balance the bubble vacuum pressure $\Delta V$.
This is possible because most SM particles become ultra-heavy inside the bubble, thereby producing an inward pressure.
For theoretical simplicity we here focus on the possibility of surrounding the bubble with a thermal bath at temperature $T$.
The system would behave as an ultra-hot star, stabilised only by artificially tuning the
temperature $T$ to be comparable to the energy difference $\Delta V^{1/4}$ between our vacuum and the unknown vacuum inside the bubble.

The value of $\Delta V$ for the possible Higgs vacuum instability is unknown.
In line of principle, $\Delta V$ could be determined (without creating the bubble)
by measuring safe few-particle collisions at futuristic ultra-high energy, and interpreting the observations in Quantum Field  Theory. 
Three extreme possibilities are: 
\begin{itemize}
\item[1)]  $\Delta V \sim M_{\rm Pl}^4$. In this case the bubble would be
uncontrollable: the true vacuum reaches the Planck scale, probing the multiverse possibly predicted by string theory.

\item[2)] $\Delta V \sim \lambda h_{\rm top}^4/4$ if the true vacuum is around $h_{\rm top}\ll M_{\rm Pl}$.
\item[3)] $\Delta V \ll h_{\rm top}^4$: this tuned possibility,
perhaps  motivated by the vague idea of Multi-Criticality~\cite{hep-ph/9511371},
would make the bubble would be more controllable.
\end{itemize}
After opening the bubble we can initially neglect it gravity, as long as
the bubble is smaller than its Schwarzschild or AdS radius, $R_{\rm s} \circa{<} M_{\rm Pl}/\Delta V^{1/2}$.
In this phase the bubble releases energy acquiring negative mass $|M| \sim \Delta V \, R^3_{\rm s} \circa{<} M_{\rm Pl}^3/\Delta V^{1/2}$.
As a numerical example, in case 2) the vacuum energy density would be $\sim 37$ orders of magnitude larger than nuclear energy, 
$R_{\rm s} \circa{<} 10^{-16}\,{\rm m}$, releasing $|M|\circa{<} 10^{11}\,{\rm kg} \sim 10^{28}\,{\rm J}$ of energy
with power $W\sim R^2_{\rm s} T^4$.
However, the bubble soon becomes an impractical source of energy:
after a time of order $R_{\rm s}$,  
preventing the bubble explosion while keeping the bubble in the weak-gravity regime costs more energy than what it released.
The bubble could be closed by `burning' it with more intense beams (corresponding to higher $T$).
Because of energy conservation, this costs at least all released energy.
Control is not established and involves dangerously fast time-scales.

\smallskip

To obtain an energy source, one needs to consider the strong-gravity regime.
In line of principle, before the vacuum bubble explodes, it is theoretically 
possible to `safely dispose' the bubble behind the horizon a pre-existing black hole.
A variant of this possibility is built in the system itself, as a black hole forms within the bubble, 
when it reaches the critical radius $R_{\rm s}$ above which its gravity is strong:
the particles remained inside the bubble undergo AdS gravitational collapse forming a black hole~\cite{Coleman:1980aw,2207.00299}.
This process does not affect the exterior.
If only vacuum energy and non-relativistic matter with density $\rho(t,r)$ is present,
the system is solved by the Lemaitre-Tolman-Bondi metric
\beq
 ds^2=- dt^2+ \frac{R'^2(t,r)}{1+2{\cal E}(r)}  dr^2+R^2(t,r)  d\Omega^2
\label{eq:TB} \eeq
where the time-independent ${\cal E}(r)$ is fixed by the initial conditions and
the location of shells with coordinate $\rt$,
$R(t,r)$, evolves as function of time as dictated by the Einstein equation
\beq
\frac{\dot{R}^2}{R^2}=\frac{1}{3\bp^2} \bigg[\frac{M(\rt)}{4\pi R^3/3}+ V_{\rm in}\bigg]+\frac{2{\cal E}}{R^2},\qquad
M'(\rt)=4\pi  R^2 R'  \rho(t,r).
\label{eq:ltb2}
\eeq
Pushing the bubble with more intense beams again should allow to close the bubble but now a small black hole remains. 
The energy balance can now be positive, as a small black hole, by itself, can be used as energy source:
by violating baryon number it converts massive particles into energy (photons and neutrinos) 
via Hawking radiation and via scatterings of gravitationally infalling particles.
A vacuum decay bubble could allow to create an artificial black hole using sub-Planckian physics.
For example, a BH with $M \sim 10^{11}\kg$ emits a power $10^{11}\,{\rm W}$
(allowing to build a 10 km-size Dyson sphere~\cite{Dyson,2106.15181} with $T\sim 300\,{\rm K}$, 
or a m-size light bulb with $T\sim T_\odot$).
It evaporates in $\sim 10^7\,{\rm yr}$, unless it's immersed in a gas with sound speed $c_s$.
Then it accretes mass within the Bondi radius $2GM/c_s^2$ with rate
$\dot M\approx 4\pi G^2M^2\rho/c_s^3$,
so gas at roughly atmospheric density balances the Hawking emission in an unstable steady state.
Since Hawking power scales as $1/M^2$ and accretion as $M^2$,
$M\approx 10^{11}\,{\rm kg}$ is roughly the lightest black hole that a gas of ordinary matter can feed.
Hard quanta with Hawking temperature $T_H \approx 100 \MeV$ do not heat too much the nearby region,
and emit charged $e^\pm$ so that the BH discharges down to $|Q| \lesssim e/\alpha$.
So it must be kept in free fall, as an electric force is much too weak for support.






\section{Conclusions}\label{concl}
We reconsidered vacuum decay stimulated by particle collisions, focusing
on the possible instability of the SM Higgs potential extrapolated up to ultra-high field values $h_{\rm top}\sim 10^{10}\GeV$.
This instability is suggested by current values of the top mass and strong coupling, 
but more accurate measurements of these quantities are needed to establish if this instability would really exist,  
and to infer a precise value of $h_{\rm top}$.

\smallskip

In section~\ref{trig} we found that Higgs vacuum decay can be triggered by an initial Higgs field configuration 
(for simplicity spherical and static)
with energy $E \circa{>} 4\pi h_{\rm top}/\sqrt{|\lambda|}\sim 500\,{\rm Joule}$
that contains a large number $N \circa{>} N_{\rm min}\sim 4\pi /|\lambda|\sim 1000$ of Higgs bosons.
Fig.\fig{dang} shows the precise threshold on the size and intensity of the initial Higgs configuration.
The critical configuration with minimal energy, the sphaleron, contains a number of Higgs quanta  mildly higher than
the critical configuration with minimal number of quanta.
Since  $N\gg 1$ these are semi-classical configurations, and
the quantum amplitude for forming a critical bubble out of collisions of few particles is exponentially suppressed by $\exp(-{\cal O}(N))$,
as typical for quantum transitions between classically different states.

This exponential suppression was more precisely computed in {\em thin-wall} approximation by Voloshin~\cite{hep-ph/9309237}.
In section~\ref{bubble} we verified this result: our result in fig.\fig{volo} confirms that particle collisions lift the exponential suppression only partially,
because the reduced exponential suppression of tunnelling gets compensated  by the exponential suppression for forming the needed semi-classical configuration.

However, the thin-wall approximation is not applicable to SM Higgs vacuum decay.
Going beyond this limit, in section~\ref{thick} we argued that
the lifting of its exponential suppression is negligible.
We provided arguments based on the approximate scale invariance of the Higgs potential, 
and assuming special field configurations that allow a simple computation.

Furthermore, in section~\ref{collider} we addressed a related issue: 
the cross section $\sigma_N$ for producing $N$ Higgs bosons grows factorially with $N$,
and some authors argue that  $\sigma_N$ might become large at  $N  \circa{>} 4\pi/\lambda $, a possibility dubbed `Higgsplosion'.
This number of quanta  (extended to  relativistic Higgs bosons) would also form the semi-classical Higgs configuration that stimulates SM Higgs vacuum decay. 
We thereby critically considered the issue, reviewing recent works that find that $\sigma_N$ remains exponentially suppressed.

\smallskip

Having clarified that stimulated vacuum decay  remains exponentially suppressed, 
in section~\ref{CR}  we compared its rate with the rate of ultra-high energy collisions of cosmic rays.
By considering collisions happened in relatively compact and possibly short-lived astrophysical production
sites of ultra-high energy cosmic rays, we estimated that the number of collisions with $\sqrt{s}\sim h_{\rm top}$
can be tens (up to 60) orders of magnitude larger than the minimal number of collisions usually estimated
in outer space, as illustrated in fig.\fig{NumberCRcollisions}.
The CR collision rate can be so high that ultra-high energy collisions among $N>2$ cosmic rays (but not $N\gg 2$) occurred.
Furthermore, the $\sqrt{s}$ of cosmic-ray collisions around their production sites could extend beyond the GZK cut-off.
Nevertheless, we find that these enhancements cannot compensate for the exponential suppression of
Higgs vacuum decay stimulated by particle collisions.

\smallskip

Finally, in section~\ref{many} we discussed how the exponential suppression can be bypassed classically
by using futuristic ultra-high energy colliders:
four or more sustained electron beams with right-handed polarization,
or a double-collision scheme producing an incoming wave of $N\circa{>}1000$ highly boosted Higgs bosons.
We wildly speculated how $\ell^-\ell^+\to h$ on-shell production could form an inward-going Higgs wave.
We simulated the classical evolution of the Higgs wave, finding the critical threshold above which
it triggers vacuum decay, rather than dissolving.
Results are shown in fig.\fig{banghik} and\fig{banglowk} for sub- and super-critical waves,
and for two different energies of the Higgs bosons.
In section~\ref{energy} we wildly speculate about a basic issue: is it possible (at least in theory) to control
a vacuum bubble slowering its expansion and using it as an energy source?
We suggest a technique based on pressing beams and on disposing the dangerous negative-mass remnant behind a black-hole horizon.


\small

\subsubsection*{Acknowledgments}
This work was supported by the MIUR grant PRIN 2017L5W2PT.
The extra work in the 2026 update was performed with the help of Claude and ChatGPT.

\end{document}